\documentclass[aps,prb,reprint,amsmath]{revtex4-2}
\usepackage{color}
\usepackage{graphicx}
\usepackage{amsfonts}
\usepackage{amsmath}

\newcommand{\zz}{\mathbb{Z}_2}

\begin{document}

\title{The gauge theory dual of the bilayer XY model with second order Josephson coupling}

\author{Pye Ton How}
\email{pthow@outlook.com}
\affiliation{Institute of Physics, Academia Sinica, Taipei 115, Taiwan}
\author{Sungkit Yip}
\email{yip@phys.sinica.edu.tw}
\affiliation{Institute of Physics, Academia Sinica, Taipei 115, Taiwan}
\affiliation{Institute of Atomic and Molecular Sciences, Academia Sinica, Taipei 115, Taiwan}

\date{\today}% It is always \today, today,
% but any date may be explicitly specified

\begin{abstract}

We formulate a duality transformation for a bilayer XY model where the layers are coupled by second order Josephson effect, which favors inter-layer phase difference of either $0$ or $\pi$.  The model may represent a bilayer superconductor or a spin-1 ferromagnetic Bose gas in the easy-plane limit.  The second order Josephson term is mapped to a U(1) gauge field, known to be trivially confining in two dimensions, and we argue that a Coulomb-gas analysis is not applicable to the dual theory.  Instead, we appeal to the vast knowledge of gauge theory and infer that the only phase transition out of low-temperature ordered phase is an Ising transition driven by condensation of $\zz$ domain wall loops. The domain wall loops can be seen as a surviving vestige of single-layer vortex-anti-vortex pair, heavily deformed by the second order Josephson coupling.  A theoretical or computational method that concentrates on point defects would most likely miss out on these excitations and reach erroneous results.  Our dual theory offers a clear, intuitive picture of how the second order Josephson coupling induces confinement of vortices and drastically changes the physics.

\end{abstract}

\maketitle

\section{Introduction}

As more experimental realizations becomes available, multicomponent superconductors and and superfluids have recently garnered renewed interests\cite{Huh2020-cy, Jiao2020-sb,  Ribak2020-dd,  Cao2021-wp, Siddiquee2022-xr,
 Zhao2023-hf, Silber2024-hm, Han2025-iy, Persky2025-bv}.  The enlarged symmetry groups of such systems generally allows for more than one possible ordered phases, with different broken symmetries\cite{Sigrist1991-mg, Stamper-Kurn2013-cz}.  Moreover, as the system is heated up from its low temperature phase, it is proposed that symmetry restoration may well occur in several steps, resulting in one or more intermediate, partially ordered phases: so-called vestigial phases\cite{Fernandes2019-yv, Hecker2023-yt}.  The proposed vestigial orders include the breaking of lattice rotational\cite{Cho2020-ne} or time reversal\cite{Hecker2023-yt} symmetries without a superconducting order, and charge-$4e$ superconductivity without breaking additional symmetry\cite{Fernandes2021-sz}.  Within the framework of Ginzburg-Landau theory, the present authors have shown that, for three dimensional (3D) systems, such vestigial orders require very extreme parameters unlikely to be realized in experiments\cite{How2023-jn, How2024-wo}.  The requirement appears more forgiving in two dimensions (2D)\cite{How2024-wh}.  Our predictions have been borne out by a recent numerical simulation \cite{Maccari2025-qf}.

In this paper, we consider a model with two complex superconducting order parameters $\Psi_{1,2}$, perhaps arising from two physical layers, gauge-invariantly coupled to each other.  In the special case where the bilinear coupling term $(\Psi_1^{*} \Psi_2 + \text{h.c.})$ is forbidden by symmetry, the leading phase-dependent inter-layer term will be the second order Josephson coupling (SOJC) $[(\Psi_1^{*})^2 (\Psi_2)^2 + \text{h.c.}]$.  Such a model can be realized in bilayer superconductors\cite{Zhao2023-hf} or a spin-1 Bose gas in the easy-plane limit\cite{Stenger1998-xq}.  Apart from the $U(1)$ symmetry that rotates both phases, this model enjoys a $\zz$ symmetry that changes the inter-layer phase difference (ILPD) by $\pi$.  At the lowest temperature, one expects quasi-long-ranged orders (qLRO) in each layer, together with a $\zz$ order that locks ILPD.  How these orders get destroyed by the rising temperature is the main question we want to address.

The SOJC term is a strongly relevant perturbation in (and possibly above) the ordered phase: a weak-coupling approximation in any treatment based in this phase is inherently questionable.  As SOJC explicitly breaks the $U(1) \times U(1)$ global symmetry of two uncoupled layers down to $U(1) \times \mathbb{Z}_2$, the spectrum of the topological excitations in the ordered phase is fundamentally altered.  Vortices in a single layer (hereby denoted as SV) are now linearly confined; only the composite vortex (two SVs, one from each layer, bound together; denoted as CV) is the sole allowed point defect.  A Berezinskii-Kosterlitz-Thouless (BKT) transition involving CV would destroy the qLRO in each layer, but there remains an long-ranged $\zz$ order in ILPD: this is the analogy of a vestigial chiral order\cite{How2023-jn, How2024-wh}.  We shall refer to this as the phase A.  On the other hand, SOJC makes possible a new type of topological excitation: the $\zz$ domain wall (DW), across which ILPD changes by $\pi$.  The proliferation of DW in, says, layer 2 destroys the qLRO in that layer only; i.e. the \emph{phase} $\phi_2 = \arg{\Psi}_2$ ceases to order, but $2\phi_2$ remains ordered.  We will refer to this as phase B.  See Table \ref{orderTable} for a summary.

\begin{table}
\caption{\label{orderTable} Properties of correlation functions of operator $\mathcal{O}$ in different phases.  Here $\phi_{1,2}$ refer to the phases of the respective layers, and for phase B the DWL condensation is assumed to occur on layer 2.  ``Y'' denotes $\zz$ order for the $e^{i(\phi_1-\phi_2)}$ and qLRO for other operators, and ``N'' denotes exponential decay of correlation.}
\begin{ruledtabular}
\begin{tabular}{l|llll}
$\mathcal{O}$ & ordered & phase A & phase B & disordered\\
\hline
$e^{i\phi_1}$ & Y & N & Y & N \\
$e^{i\phi_2}$ & Y & N & N & N\\
$e^{i(\phi_1 - \phi_2)}$ & Y & Y & N & N\\
$e^{2i\phi_2}$ & Y & N & Y & N
\end{tabular}

\end{ruledtabular}
\end{table}

The XY model in 2D, along with any number of related models in a big, extended family, have seen decades of studies; see \cite{Korshunov2006-zl} for a relatively recent review.  Bilayer XY model with first order inter-layer Josephson coupling has been studied \cite{Parga1980-gb,Mathey2008-ff,Bighin2019-xp}.  Closely related to our case is the extended XY-model with competing periodicities\cite{Lee1985-qy,  Granato1986-dc, Granato1988-nv, Kobayashi2020-vu}, which shares a similar $\mathbb{Z}_p$ structure (our case has $p = 2$).  The model also features domain walls and fractional vortices.  Despite the similarities, the fact that it has only one layer as oppose to two leads to some notable differences in the duality transformation.  Also closely related and well-studied is the coupled Ising-XY model\cite{Choi1985-qn, Granato1991-er, Li1994-cm}.  The model studied in this papers has been previously considered in \cite{Granato1986-dc, Granato1988-nv, Song2022-dn, Liu2023-ce,  Zeng2024-qg}, and we will compare our result with the prior literature as we go.  Also see \cite{Bonnes2012-vw, Maccari2023-cs}, where the model is supplemented by extra constraints.  A model effectively bilayer with an even higher order Josephson coupling, motivated by the $\mathbb{Z}_3$ nematicity in trigonal lattices, has also been studied\cite{Jian2021-oe}.  Along a different direction, the bilayer model coupled not with a Josephson term but a non-dissipative drag term is studied in \cite{Karle2019-rq}.

Duality is an extremely helpful tool for treating strong-coupling physics, mapping high temperature to low, particle to topological excitation, order to disorder, and strong coupling to weak \cite{Jose1977-kb, Savit1980-kt, Dasgupta1981-nq, Lee1991-ye, Seiberg2016-rb, Ma2018-zd}.  The XY model in 3D is dual to the U(1) Abelian Higgs model (AHm) \cite{Peskin1978-qq}.  Considering the bilayer problem as a 3D model with the $z$-direction truncated to only two lattice sites, the $z$-component of the dual gauge field would become another set of ``matter'' fields, and the remaining two in-plane components form a 2D U(1) gauge theory.  Guided by this intuition, our duality transformation results in a generalized 2D AHm with multiple matter fields of different charges, representing vortices and domain walls in the original model.  

The U(1) gauge theory in 2D coupled to scalar matters has been very well studied \cite{Callan1977-lc, Witten1979-lm, Coleman1985-tv, Komargodski2019-sp}, and the duality mapping provides a clear, well-understood language to the physics that is otherwise more convoluted.  The gauge field is trivially confining, and forbids any configuration with non-vanishing charge.  The theory governing the topological excitations thus bear no resemblance at all to the Coulomb gas picture.  We are able to argue that the $\zz$ Ising transition, driven by closed loops of domain walls (DWL), is the only realistic possibility when the two layers are not identical, barring extreme fine-tuning of the material property.  The DWL can be seen as a surviving, heavily-deformed vestige of a single-layer vortex-antivortex pair, which has no out-of-plane vorticity anywhere, but nonetheless is a topologically protected \emph{line} defect.  This Ising transition is therefore a continuation of the single-layer BKT transition in the decoupled limit with zero SOJC.  We also conjecture that, for identical layers, a XY-class transition takes place.

This paper is organized as follows.  We start by reviewing some expected consequence of SOJC, and how topological excitations behave in its presence.  The dual lattice model is then constructed, and we analyze its effective continuum field theory to extract the phase diagram, appealing to the knowledge of AHm in 2D.  We comment on the addition of superfluid drag.  We finally compare our result with existing literature and make our conclusion.

\section{Implications of SOJC}

Let us first write down the model to be considered.  We consider the polar representation of the order parameters $\Psi_i = \vert\Psi_i\vert e^{i\phi_i}$, and ignore amplitude fluctuations, leaving only the phases $\phi_i$ as the dynamical variables.  The system is placed on a (bilayer) square lattice to allow for topological excitations.  Our starting point is then two copies of square lattice XY model with SOJC:
\begin{equation}
\label{XY}
S\! = \!\!\sum_{\substack{i=1,2 \\ r}}\!\!\! -K_i \cos(\vec{\Delta} \phi_i)  - \sum_{r} \lambda \cos(2\phi_1 - 2\phi_2).
\end{equation}
Here $r$ labels the 2D lattice sites, and $\cos(\vec{v})$ is a shorthand for $\cos(v_x) + \cos(v_y)$

\subsection{Failure of the Coulomb gas formalism}

The Coulomb gas formalism is a powerful theoretical tool for problems related to sine-Gordon theory and XY model in 2D.  In particular, by treating vortices as a Coulomb gas, the formalism underpins the renormalization group (RG) analysis of the BKT transition \cite{Kosterlitz1974-ih, Amit1980-nc}.  Look another way, it is in essence a weak-coupling expansion around the free boson conformal fixed point.  We will, however, presently show that this expansion is inappropriate for our model \eqref{XY}.

Consider \eqref{XY} in the low temperature ordered phase, where topological excitations can be ignored as a first approximation.  It is then appropriate to take instead the continuum effective field theory
\begin{equation}
\label{XYcontinuum}
\mathcal{L} = \sum_{i = 1,2} \frac{K_i'}{2} \vert\nabla \phi_i \vert^2 - \lambda' \cos(2\phi_1 - 2\phi_2)
\end{equation}
with only smooth field configurations.

Around $\lambda' = 0$, the scaling dimension of the SOJC operator is $2 - \frac{1}{\pi}\left(\frac{1}{K_1'} + \frac{1}{K_2'}\right)$.  On the other hand, exactly at $\lambda' = 0$, the single-layer BKT transitions occur at $K_i' = \frac{2}{\pi}$.  If one performs perturbation around the decoupled limit $\lambda' = 0$ and comes down from high temperature, when both layers are separately ordered in the decoupled limit, any SOJC perturbation has long since become strongly relevant.  Consequently, any approach that treats the SOJC term as a weak perturbation is of questionable validity: it should be either completely absent or strongly coupled near any possible phase transition.  In their early work on bilayer XY model with an ordinary Josephson coupling, Parga and van Hinbergen already noted the same difficulty and concluded that the interacting Coulomb gas picture is unreliable there\cite{Parga1980-gb}.

\subsection{Topological excitations}

In the low temperature ordered phase, consider the hypothetical configuration of a cylindrical-symmetric SV in one layer over the other uniform layer.  It is easy to see that the region where ILPD significantly deviates from zero or $\pi$ scales as \emph{total area} of the system, and it becomes prohibitively costly to create a single SV with SOJC turned on.  To avoid this energy cost, one can binds two SVs, one from each layer, so that the phases winds together and the ILPD is zero (or $\pi$) everywhere.  The resultant CV configuration has a cost logarithmic in system size, and behaves very much like a vortex in the single-layer problem.  It is not a stretch to imagine a BKT transition driven by CV condensation, sending the system from the ordered phase to phase A.

\begin{figure}
\includegraphics[width=0.5\textwidth]{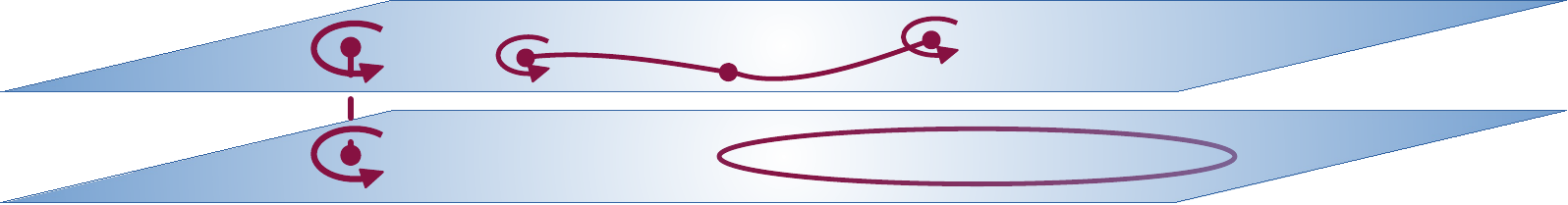}
\caption{\label{topoExcite} Topological excitations of the bilayer model.  In the illustration are a CV involving both layers, an SV stretched out into two HVs and the domain wall linking them in the top layer, and a DWL in the bottom layer.  Note that the configuration of a DW with two HV end points is not a truly stable, as discussed in the text.}
\end{figure}

One may also try to deform an SV, stretching out its core into a long defect line (thickness of the order of healing length).  Locally, on either side of the line the phase appears uniform, but there is a $\pi$ discontinuity across the line, and this is identified as the $\zz$ DW.  The vorticity of the starting SV is split into two half-vortices (HVs) of the same sign, now situated at the ends of the DW.  We will view the SVs and HVs as having vorticity in the $z$-direction, and the DW lines themselves as \emph{in-plane half vortex lines}\cite{Lee1985-qy, Gao2022-ry}.  Being a defect in superfluid density, DW sustains a tension, but it is energetically favorable for an SV to be deformed this way.  Consider a straight DW of length $2R$ with HVs attached to the ends.  Within an area of $R^2$, the phase remains roughly uniform and the DW tension (the only energy cost) scales as $R$; beyond the $R^2$ area, the configuration is just a deformed SV and the cost again scales as area.  It is therefore energetically favorable for an SV to split.  But to have a configuration with truly finite energy, one can only start with a pair of SV and anti-SV, and deform them into a closed DWL with \emph{zero net vorticity}.  The condensation of DWLs is another possible phase transition out of the ordered phase, resulting in phase B.  Another way to view phase B is that only one layer is truly (quasi-)ordered, and the other layer gets an induced order by proximity effect through the SOJC.  See Fig \ref{topoExcite} for illustrations of the topological configurations discussed above.

The usual gradient energy and the SOJC separately contribute to the cost of vorticity.  While the SOJC cost is avoided by matching the vorticities of the two layers, the gradient term still imposes the familiar condition of vanishing net vorticity within a single layer.  Crucially, SOJC does not facilitate the interconversion of vorticities between the layers.  For example, a CV has one unit of vorticity in each layer, and must always be paired with an (anti)-CV to ensure the net-zero-vorticity condition is met.  By the same token, a DW line must remains within the same layer.  If a DW line ever jumps from one layer to the other, it leaves a dangling HV on each layer at the point of jumping, breaking the net-zero-vorticity condition.  If we are to have a dual description that treats these topological defects as fundamental ingredients, it should contain two sets of separately conserved topological charges, one for each layer.

%
%
%
%
%
%One thing is clear from this discussion: vorticity cannot jump from one layer to the other.  While the binding of two SVs into a CV relieves the system from the SOJC energy cost, each layer still sees one unit of vorticity and a logarithmically diverging cost; a CV does not have zero net-vorticity and must always be paired.  The DW, considered as in-plane HV line, also has to stay within a layer: a DW line jumping from one layer to the other leaves behind a pair of dangling HVs, one on each layer, and energetically costly.  A successful dual description of the model must therefore contain two sets of separately conserved topological charges, one for each layer.

Also note that SOJC is an explicit symmetry-breaking term: the symmetry of the bilayer model is $U(1)\times U(1)$ without it and $U(1) \times \mathbb{Z}_2$ with it.  Standard homotopy theory immediately tells us that the CV is the only topologically stable point defect in the presence of SOJC.  But pair of SVs with opposite vorticities lives on, being deformed into DWL.  Any theoretical treatment focusing solely on point defects would likely erroneously conclude that SVs are tightly confined and not relevant to the melting of the low-temperature order, while missing out the DWL-driven transition.

\section{Duality transformation}

We begin by recalling the duality transformation of the XY model in 3D \cite{Peskin1978-qq}.  Starting with one scalar field on the XY-model side, the dual theory turns out to be a compact scalar field coupled to a non-compact U(1) gauge field, i.e. the AHm.  Vorticity is the dual gauge charge.  The conserved current becomes the dual magnetic field, and the vortex line is identified with the \emph{world line} of the dual matter excitation \cite{Peskin1978-qq, Lee1991-ye, Gao2022-ry}.  A vortex line starts and ends on XY-solitons, which are created and annihilated by the dual particle operators.

We write down the equivalent Villain model\cite{Villain1975-ed} of \eqref{XY}, with a so-called \emph{fictitious layer} inserted between the two physical layers:
\begin{widetext}
\begin{equation}
\label{Villain}
S\! = \sum_{\substack{i=1,2 \\ r}} \! \frac{K_i}{2} \left( \vec{\Delta} \phi_i + 2\pi \vec{m}_i\right)^2
+ \sum_{r} \! \frac{K'}{2} \left( \vec{\Delta} \phi' + 2\pi \vec{m}' \right)^2
+ \sum_{r} \lambda\left[ (2\phi_1 - 2\phi' + 2\pi m_{z1})^2 + (2\phi' - 2\phi_2 + 2\pi m_{z2})^2 \right]
\end{equation}
\end{widetext}
The limit $K' \rightarrow 0^{+}$ is implied.  With $K'$ set to zero, the fictitious $\phi'$ can be exactly integrated out to recover the naive two-layer Villain model.  This is seen as a mini 3D model, with the $z$-direction truncated to only three layers.

The rest of the duality transformation follows Peskin\cite{Peskin1978-qq}, with suitable changes accommodating for the $\pi$-periodicity of the SOJC term, and the truncation in the $z$-direction\footnote{See Supplemental Material.}.  A direct consequence is that only integer ($2\pi$) vorticity can pierce through a layer, and only half vortex lines ($\pi$) criss-cross the space between layers.  Following previous discussion, these half vortex lines are identified as DWs.  Also as previously argued, the vorticity of the two physical layers should be separately conserved.  We enforce the condition by setting an \emph{infinite} cost for vorticity going through the fictitious layer during the duality steps.  The dual lattice model takes the final form:
\begin{widetext}
\begin{equation}
\label{dual_lattice}
S_d \! = \sum_r \frac{(\vec{\Delta} \times \vec{a})^2}{8\pi^2\lambda}
	+ \sum_{\substack{i = 1,2 \\ r}} 
			- \frac{\cos(\vec{\Delta}\theta_i - (-1)^i 2\vec{a})}{4\pi^2 K_i}
			- \frac{\cos(\vec{\Delta}\chi_i - \vec{a})}{V_i}
			- \frac{\cos(2\chi_i - (-1)^i \theta_i)}{U_i}
\end{equation}
\end{widetext}
Here $\vec{a}$ is a non-compact U(1) gauge field in 2D, and $\theta_{1,2}$ and $\chi_{1,2}$ are compact (angular) variables in the dual theory, where the subscript refers to the layers.  During the intermediate steps one must insert by hand the costs of various topological defects \cite{Peskin1978-qq} (also see Appendix B); these are $U_{1,2}$ the SV chemical potentials, and $V_{1,2}$ the DW tensions of their respective layers.

Let's describe the duality dictionary.  The dual magnetic field $\vec{\Delta}\times \vec{a}$ is the interlayer tunnelling current due to SOJC.  Thus any SOJC physics is mapped to the gauge sector in the dual theory.  An SV in layer $i$ (sans the tunnelling current) is created by $e^{\pm i \theta_i}$, and the in-plane physical current in layer $i$ is dual to the (gauge-invariant) topological current of $\theta_i$.  As in the 3D case, vorticity becomes gauge charge in the dual theory.  (The gauge field $\vec{a}$ being non-compact, we are free to rescale it so that one unit of vorticity is dual to \emph{two} units of gauge charge.)

From its coupling to the gauge field, $\chi_i$ is seen to have a gauge charge that is \emph{half} the magnitude of that of $\theta_i$.  Therefore $e^{\pm i \chi_i}$ creates a HV (sans tunnelling current) in layer $i$, which is the dangling end point of a DW.  The \emph{world line} of such a HV excitation is then identified as the DW line itself.  In the spirit of particle-vortex duality\cite{Lee1985-qy}, the (gauge-invariant) $\chi_i$ current on the dual side furnishes a coarse-grained description of the DW lines in layer $i$.

The gauge invariance principle of the dual theory has a very simple interpretation: any physical configuration must come with its associated tunnelling current.  Thus, the operators $e^{\pm i \theta_i}$ and $e^{\pm i \chi_i}$ that insert ``naked'' topological charges are not physical by themselves; the excitation must be dressed by its associated gauge field.

\begin{figure}
\includegraphics[width=0.45\textwidth]{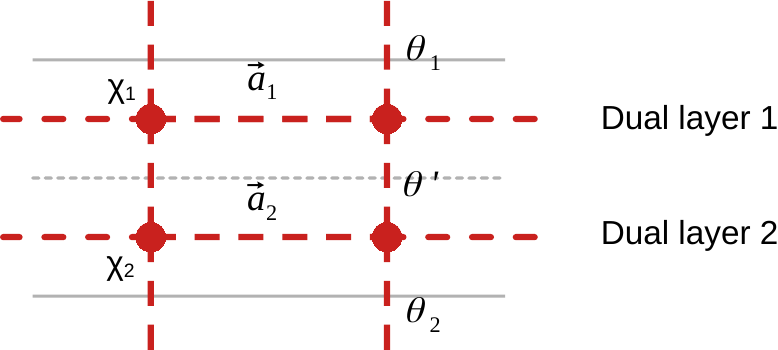}
\caption{\label{dualFigure} A cross section view of the dual lattice.  The gray solid and dashed lines indicate the original bilayer and the fictitious layers, respectively.  The dual gauge field (defined on dual links) in 3D decomposes into 2D gauge field $\vec{a}_{1,2}$ and the $z$-components $\theta_{1}, \theta_{2}$, and $\theta'$.  The fictitious layer condition forces $\vec{a}_1 = \vec{a}_2$, and one can make a partial gauge choice to fix $\theta' = 0$.}
\end{figure}

It is instructive to compare \eqref{dual_lattice} with what one would naively expect from truncating a 3D model.  In the full 3D case, the center of a cube becomes a dual lattice site, where the dual matter field lives; a plaquette becomes a dual link between to dual sites, on which the dual gauge field is defined.  The dual of our three-layer Villain model \eqref{Villain} should therefore consists of: two layers of dual matter field and in-plane gauge field, and three sets of $z$-component gauge field (above, between and below the two dual layers).  See Fig \ref{dualFigure}.  The dual matter field are identified with $\chi_{1,2}$.  We are free to make a partial gauge choice that set the $z$-component gauge field to zero in between the dual layers, and the remaining $z$-component gauge fields (above and below) are rebranded into charge-$\pm 2$ dual matter fields $\theta_{1,2}$.  The limit $K' \rightarrow 0^{+}$ forces the two sets of in-plane gauge field to be equal, so there is only one set of 2D U(1) gauge field $\vec{a}$.  Finally, the infinite cost of vorticity through the fictitious layer forbids any direct coupling between the two dual layers.  When the dust settled, one can see that the first two terms of \eqref{dual_lattice} come from the 3D Maxwell term, and the last two terms come from 3D covariant derivatives of the $\chi$ field.

%
%Notably, the SOJC term is dual to the Maxwell action.  Any peculiarity of the dual gauge theory traces its root to the presence of SOJC.
%
%One can also read off the duality dictionary from the comparison.  The conserved XY current is identified with the dual magnetic field in the full 3D case.  The 2D dual magnetic field $\vec{\Delta}\times\vec{a}$ therefore represents the \emph{interlayer} tunneling current, and the topological currents of $\theta$ represents the \emph{in-plane} current.  Furthermore, the decoupled limit is obtained by setting $\lambda$ to zero, which forces $\vec{a}$ to be gauge-trivial.  If one ignore the $\chi_i$ fields that represent the half vortice and have no reason to be there in the decoupled limit, the remaining $\theta_i$ fields coincides exactly with the disorder field obtained from duality of single-layer XY model.  The operators $e^{\pm i \theta_{1,2}}$ create SVs \emph{sans} inter-layer tunneling current in their respective layers.
%
%Similarly, the operators $e^{\pm i \chi_{1,2}}$ create HVs without tunneling current.  The in-plane 2D world lines of $\chi$-excitations are identified with DW lines, a correspondence directly inherited from the full 3D case.  And we see the identification of winding numbers and dual gauge charges: $\theta_i$ has twice the charge of $\chi_i$.
%
%Note that these HV and SV configurations above are incomplete and unphysical in the sense that no inter-layer tunneling current is created alongside.  In the dual theory this is reflected by the operators not being gauge invariant.
%
Rewrite \eqref{dual_lattice} using vorticity operators $H_i \sim e^{i \theta_i}$ and $\psi_i \sim e^{i\chi_i}$.  Softening their amplitudes, and one obtains an Euclidean field theory whose critical behavior is dual to the original bilayer superconductor\cite{Dasgupta1981-nq}:
\begin{widetext}
\begin{equation}
\label{FT}
\mathcal{L} = \frac{(\nabla \times \vec{a})^2}{8\pi^2\lambda} + \sum_{i= 1,2}
	\left(\vert \vec{D} \psi_i \vert^2 + \vert \vec{D} H_i \vert^2
	+ m_i \vert\psi_i\vert^2 + \mu_i \vert H_i\vert^2
	+ u_i \vert H_i\vert^4 + v_i \vert \psi_i \vert^4 \right)
	+ g_1 \left(H_1 \psi_1^2 + \text{h.c}\right) + g_2 \left(H_2^{*} \psi_2^2 + \text{h.c}\right).
\end{equation}
\end{widetext}
Here $\vec{D} \equiv \vec{\nabla} - q \vec{a}$ is the covariant derivative.  The charges $q$ for the fields are: $+1$ for both $\psi_{1,2}$, $-2$ for $H_1$, and $+2$ for $H_2$.  Apart from the local U(1) gauge transformation, there are two separate \emph{global} U(1) symmetries, each confined to one layer.%
\footnote{Reader familiar with the duality of the single-layer model may recall that the dual U(1) symmetry is explicitly broken by the vortex fugacity.  Similar terms do exist here, but they gain the gauge-invariant form $\cos(\theta_i \pm 2 \chi_i)$.  See also Ch 25 of \cite{Sachdev2023-fd}.}  To put it the other way round, the gauge redundancy introduced into the dual theory forbids any explicit breaking of the dual U(1) symmetry, as per Elitzur's theorem.
This is the direct consequence of the separate vorticity conservation within each layer.  Other terms not forbidden by symmetry can and should be included, but they bear no consequence on our subsequent analysis.

\section{Critical Theory}

We will start our analysis from the dual disordered phase $m_1, m_2, \mu_1, \mu_2 \gg 0$, which is the original XY-ordered phase.  The fields $H_{1,2}$ create SV excitations in their respective layers.  Apart from being gapped, these are, of course, linearly confined by the U(1) gauge field in 2D and never appear in the physical spectrum of the theory\cite{Komargodski2019-sp}.  The dual-gauge-invariant bound states are generally gapped\cite{Jones1979-zo}.  The fields $\psi_{1,2}$ create what are half vortices in the original model, and serve as sources of DWs.  Due to gauge confinement, only the \emph{Wilson loops} are physically allowed and these are identified with DWLs.  The gaps $m_{1,2}$ are DW tensions.  The Yukawa-like $H_1 \psi_1^2$ and related terms describe the process of an SV stretching into a long DW.

Notably, there are no cross terms like $H_1 H_2$ or $\psi_1^{*} \psi_2$, despite them being gauge invariant.  Their absence is rigidly protected by the two separate global U(1) symmetries.  We identify $H_1 H_2$ as the fugacity term of CV, and its absence lead to our first conclusion: \emph{a BKT-type transition out of the XY-order phase, driven by CV proliferation, is not possible.  Phase A is forbidden.}

Indeed, if we ``switch off'' the $\psi_{1,2}$ fields, leaving the matter solely in the hand of $H_{1,2}$, there would be no phase transition at all.  It is known that AHm in 2D does not have a Higgs phase: the would-be qLRO is always destroyed by vortex-instantons\cite{Callan1977-lc}.  Not to be confused with the original XY model vortices, these are solitonic configurations of the dual fields $H_{1,2}$: ``vortex of vortex'', so to speak, and are dual to amplitude fluctuations in the original model.

The alerted reader may raise an objection: surely this dual vortex breaks the U(1) symmetry of the original model and therefore cannot be allowed\cite{Wen2007-xf}?  For the single-layer case, a dual vortex is a source of the in-plane XY current, and is indeed forbidden.  But this is not a concern here: the current simply tunnels from the other layer.  Indeed, formally the familiar Abrikosov vortex solution applies here, and the total dual magnetic flux threading the vortex is locked to the dual winding number.  Translating back to the original XY model, the total tunnelling current equals the radial in-plane current, and current conservation is not violated.

Let's put back the $\psi$ fields.  Assuming the two layers are not identical, and layer two is the weaker superfluid, one expects the critical behavior of the model to be dictated by layer two only.  As the physical temperature is raised, there must indeed be a point where $m_2 < 0$ and $m_1 > 0$ in the dual theory.  Even though $H_2$ cannot really order, the Yukawa-like term $g_2 (H_2^{*} \psi_2^2 + \text{h.c.})$ Higgses $\psi_2$ into a \emph{real} field.

To be precise, let us introduce the polar representations $\psi_2 = \rho_2 e^{i\chi_2}$ and $H_2 = \vert H_2 \vert e^{i\theta_2}$.  Away from isolated points of $H_2$ vortex centers in space, $\nabla \chi_2 = \vec{a} = -\nabla \theta_2/2$, though discontinuity of $\pi$ in $\chi_2$ itself is allowed, making $\rho_2$ possibly negative (and thus spanning the entire real line).  The covariant derivative term $\vert \vec{D} \psi_2 \vert^2$ becomes simply $\vert \nabla \rho_2\vert^2$, and $\rho_2$ is decoupled from the gauge field.  With correct phase alignment, the Yukawa term is made into $-2 \vert g_2 \vert \vert H_2\vert \rho_2^2$ and pushes $\rho_2$ toward ordering.  In effect one obtains:
\begin{equation}
\label{Ising}
\mathcal{L} = \dots + m_2' \rho_2^2 + u' \rho_2^4
\end{equation}
which describes the condensation of the real field $\rho_2$, an \emph{Ising transition}.

The Higgsing and decoupling from gauge field is only asymptotic at large distance, which is an appropriate limit for critical behavior.  Away from the critical point, $\psi_2$ is still a charged field and linearly confined, and the Ising transition really describes the condensation of gauge-invariant Wilson loops.  We therefore reach our second conclusion: \emph{when the two layers are non-identical, the system undergoes DWL condensation in the weaker layer, which is an Ising transition.  Phase B is always above the XY-ordered phase.}

We emphasize that a DWL is essentially made out of a pair of deformed SVs of opposite sign: there is no out-of-plane vorticity due to the confinement effect of SOJC, but the defect remains.  In the presence of SOJC, the global symmetry is lowered to U(1)$\times \mathbb{Z}_2$ and, following the usual homotopy argument, SV is not even a legitimate point defect.  However, DWL is a surviving vestige of the SV-sector from the decoupled theory.  The DWL-driven transition is therefore a continuation of single-layer BKT transition when the layers are decoupled.  This is an important intuition that we will revisit when making comparison with the sine-Gordon calculation\cite{Liu2023-ce, Zeng2024-qg}.

What if the two layers are identical?  Recall that $\psi_{1,2}$ are similarly charged and strongly repel each other through gauge interaction.  When they are Higgsed into the real $\rho_{1,2}$, one would expect a residual repulsion $\rho_1^2 \rho_2^2$ term with a positive coefficient.  If $\epsilon$-expansion from four dimensions is still credible at all, one expects the renormalization group (RG) flow to asymptotically restore the O(2) symmetry among $\rho_{1,2}$\cite{Patashinskii1979-hr}.  Our third conclusion is more of an conjecture: \emph{when the two layers are identical, there is a single phase transition in the XY-class (BKT-type) that destroys phase coherence in both layers.}

As the duality transformation essentially assume diluteness of all vorticity\cite{Jose1977-kb, Peskin1978-qq}, we deem the dual theory only appropriate for the investigation of the first transition out of the low-temperature XY-ordered phase.  However, phase B is easily seen as having qLRO in one layer, and some partial order \emph{by proximity} in the other.  It must be separated from the disordered phase by a single-layer BKT transition.  Our proposed phase diagram is thus identical to that in \cite{Song2022-dn}, though our interpretation of the Ising transition differs.

\begin{figure}
\includegraphics[width=0.4\textwidth]{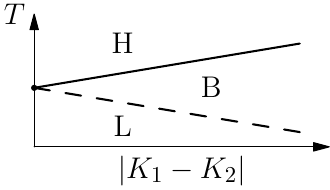}
\caption{\label{phaseDiagram} The phase diagram.  H denotes the high-temperature disordered phase, and L the low-temperature ordered phase.  The B phase is defined in the text.  The solid line is the single-layer BKT transition, and the dashed line is the Ising transition.  When the layers are identical, the only transition is also of BKT type.}
\end{figure}

\section{Superfluid drag}

An inter-layer current-current coupling is entirely consistent with the symmetry of the bilayer problem, and induces the so-called superfluid drag effect: the supercurrent in one layer stirs up a parallel (or anti-parallel) response supercurrent in the other\cite{Andreev1975-tr}.  Notably, when anti-parallel response is favored, the system may enter the \emph{counterflow superfluid} phase, which recently sees experimental realization\cite{Zheng2025-zy}.  The phase diagram in the presence of the drag term has been studied extensively, though the drag is often regarded as the most dominant inter-layer coupling, and any Josephson term is inapplicable to the underlying physical system (e.g. BEC of two species) \cite{Kaurov2005-oi, Fil2005-sm, Dahl2008-wz, Bighin2019-xp, Karle2019-rq, Underwood2023-ao}.

The inclusion of a superfluid drag adds the term $K_d \nabla\phi_1 \cdot \nabla \phi_2$ to the continuum model \eqref{XYcontinuum}.  This term in general is not amenable to a lattice XY model formulation such as \eqref{XY}, but one can infer that a positive (negative) $K_d$ favors binding of SVs with opposite (equal) vorticities from the two layers, among other effects it can induce.  We will refer to this composite object as a $(1,-1)$-vortex ($(1,1)$-vortex).  Note that the $(1,1)$-vortex is just another alias for CV introduced earlier, but the $(1,-1)$-vortex runs afoul with confinement if SOJC is present.

Let us take a step back and consider a bilayer system with superfluid drag but \emph{no Josephson coupling}, modelled by the simple continuum Hamiltonian density
\begin{equation}
\begin{split}
\mathcal{H} &= \frac{K}{2}\left(\vert\nabla \phi_1\vert^2 + \vert\nabla \phi_2\vert^2\right) + K_d \left(\nabla \phi_1\right)\cdot\left( \nabla \phi_2\right)\\
&= \left(\frac{K+K_d}{4}\right) \vert \nabla \phi_+\vert^2 + \left(\frac{K-K_d}{4}\right) \vert \nabla \phi_-\vert^2,
\end{split}
\label{dragModel}
\end{equation}
where $\phi_{\pm} \equiv \phi_1 \pm \phi_2$.  We make the simplifying assumption of two layers being identical here.  Written in the suggestive form of the second line, one is lead to wonder if, as the temperature is raised, the system would first undergo a BKT transition in one of the $\phi_{\pm}$ components instead.

Standard estimation for the SV-driven BKT transition temperature in a single layer is roughly $T_{\text{SV}} \sim K$, omitting a multiplicative factor.  On the other hand, the $(1,\pm 1)$-vortex introduced above corresponds to $4\pi$ vorticity in $\phi_{\pm}$, respectively.  The same estimation for their BKT temperature yields $T_{(1,\pm 1)} \sim 2(K\pm K_d)$, up to the same multiplicative factor.  We thus arrive at an estimation: when $K \pm K_d < K/2$, the proliferation of $(1,\pm 1)$-vortex pre-empts the transition driven by SV as the temperature is raised, and the system enters a phase where only $\phi_{\mp}$ retains a qLRO.  Ordering of only $\phi_{+}$ can be interpreted as a charge-4e superconducting phase, and ordering of only $\phi_{-}$ a vestigial chiral phase.

Crucially, this argument provides that the size of $K_d$ must be comparable to $K/2$ to force these new phases.  Explicit and intentional engineering toward the limit is of course possible\cite{Zheng2025-zy}, but we believe this $K_d$ is unrealistically large for materials not fine-tuned for it.  As we modify \eqref{dragModel} by adding SOJC and making the layers asymmetric, we expect the phase diagram Fig \ref{phaseDiagram} \cite{Song2022-dn} to qualitatively hold for realistic values of $K_d$.

\subsection{comparison to existing sine-Gordon calculations}

With a fixed SOJC and varying drag, we do anticipate a phase diagram qualitatively similar to that presented in \cite{Liu2023-ce, Zeng2024-qg}.  However, we disagree with the critical strength of the drag obtained in both works as being too small.

Ref \cite{Zeng2024-qg} does not have a drag term in their initial lattice Hamiltonian, but by allowing the ratios $K_+/K_-$ and $g_{\phi_+}/g_{\phi_-}$ in their study to be arbitrary, the drag effect is implicitly present in their sine-Gordon model.  From their subsequent treatment, it is clear that \emph{all} phase transitions obtained there are principally driven by the terms that create $(1,\pm 1)$-vortices in our own language.

Drag is explicitly present in the initial Hamiltonian in \cite{Liu2023-ce}, and they arrive at a sine-Gordon model essentially identical to that in \cite{Zeng2024-qg}.  The threshold values of what is our $K_d$ can be read off their Fig 2b: $K_d > K/3$ or $K_d < -K/5$ for the presence of the vestigial phases.  In this work there is also a Monte Carlo study not based on sine-Gordon analysis.  In the supplemental material, their Fig S5 shows the Monte Carlo phase diagram obtained without the so-called ``kinetic constraint'', and both transition lines can be clearly identified with the BKT transition of $(1,\pm 1)$-vortices.  The ``kinetic constraint'' only serves to deform these phase boundaries.

DWL is completely absent in either studies.  Our duality transformation highlights the importance of DWL, which could be seen as a vestige of SVs that are confined by a non-zero SOJC.  As such, when the SOJC is continuously tuned toward zero, the DWL-driven transition should connect to the single-SV BKT transition.  Following our preceding discussion, by continuity, we expect the DWL-driven transition to pre-empts the $(1,\pm 1)$-vortex BKT transition in a large region of the parameter space, until the drag gets sufficiently large, $\vert K_d\vert \approx K/2$.  Again, barring deliberate fine-tuning, we believe this is rather extreme and unlikely in realistic material.

\section{further Comparison to literature}

The earliest result on this bilayer model to our knowledge is \cite{Granato1986-dc, Granato1988-nv}.  While their proposed phase diagram (Fig. 2 there) is compatible with ours, the validity of the weak-coupling RG analysis, treating SVs as a Coulomb gas, is questionable.  Their strong coupling analysis shows linear confinement and stretching of SVs in a more convoluted way, but the numerical RG calculation in that case is not conclusive.

Tensor network method is employed to numerically simulate the bilayer model in \cite{Song2022-dn}, and the result is fully compatible with our analysis.  The only disagreement is in the interpretation: they interpret the Ising transition as dissociation and proliferation of HV pairs joined by DW lines, while we believe there should be no dangling HVs, and DWs always form closed loops.  Our duality treatment provides a transparent and intuitive picture for their numerical results.

The related problem of a two-component \emph{nematic} superconductor has also been studied in the \cite{Jian2021-oe}.  Our duality transformation cannot directly apply to the different model without appropriate modification, but based on what we have learned here, the three types of domain walls between the three degenerate nematic states should play an important role in the nematic model, too.  DW is not present in the 2D sine-Gordon analysis of \cite{Jian2021-oe}, and we suspect that they overestimated the transition temperature by only allowing BKT transition driven by what's the analogy of our CV.

We want to mention the hexatic-nematic liquid crystal model studied in ref \cite{Drouin-Touchette2022-fr}.  The model is further removed from our present problem, but we note a shared common theme: extended topological excitations, formed by deforming and fractionalizing point-like vortex, turns out to drive a low-temperature phase transition.  The physics would've been missed if only point-like excitations were allowed in the theoretical treatment.

There are also attempts to directly simulate related models using Monte Carlo methods.  However, they both comes with additional constraints that the respective authors believe better reflects the physics.  Ref \cite{Bonnes2012-vw} that studies a Bose Hubbard model applies a constraint that forbids triple on-site occupancy, while Ref \cite{Maccari2023-cs} allows density fluctuations of each component but imposes the condition that total superfluid density is constant.  We will refrain from a direct comparison due to these differences.

\section{Conclusion}

We study a model where two copies of 2D XY models are coupled together by a SOJC term.  The inclusion of SOJC completely changes the behavior of the model, leading to linear confinement of SVs and the emergence of $\zz$ DW as topological excitations.  Motivated by the observation that the bilayer model is an extreme truncation of a 3D model in one direction, and that the XY model in 3D is known to dual to a U(1) gauge theory, we formulate a duality transformation that maps the bilayer model to a generalized AHm in 2D: the dual matter fields represents SVs and DWs, while the dual U(1) magnetic field is the tunnelling current due to the SOJC.

Our duality transformation provides a concrete physical picture of the ordered phase of the bilayer model, clearly revealing topological excitations as the dual degrees of freedom.  The emergent U(1) gauge structure is well-understood, and we are able to deduce many non-perturbative properties of the model appealing to our knowledge of U(1) gauge theory in 2D.

The U(1) gauge structure immediately implies linear confinement of vorticity.  We are also able to rule out any transition out of the XY-ordered phase driven by integer vortices.  When two layers are not identical, as the physical temperature is raised, we generally expect the proliferation of DWLs to destroy phase coherence in the layer with weaker superfluidity.  We identify the transition to be in the Ising class.  When they layers are identical, we conjecture that the RG flow asymptotically restore an inter-layer O(2) symmetry, and a single XY-class transition destroy phase coherence in both layers.

\appendix

\section{Renormalization of the second order Josephson coupling}

We will work out the one-loop beta function of the continuum field theory model, treating the SOJC perturbatively.  The continuum 2D Euclidean field theory to be considered is
\begin{equation}
\begin{split}
\mathcal{L}_0 &= \frac{J_1}{2}\vert\nabla\phi_1\vert^2 + \frac{J_2}{2}\vert\nabla\phi_2\vert^2\\
\delta\mathcal{L} &= \sum_{i=1}^{3} g_i \varphi_i%.
\end{split}
\end{equation}
We consider $\mathcal{L}_0$ a fixed point Lagrangian, to be deformed by the operators $\varphi_1 = \vert\nabla\phi_1\vert^2$, $\varphi_2 = \vert\nabla\phi_2\vert^2$ and $\varphi_3 = \cos(2\phi_1 - 2\phi_2)$ in $\delta\mathcal{L}$.

Assuming the system is deep in the ordered phase at a low temperature (that is, $J_1$ and $J_2$ are large enough), we can ignore topological excitations in the subsequent calculation.  We follow the renormalization procedure first described by Cardy \cite{Cardy1996-ml}.  Relative to $\mathcal{L}_0$, let $\Delta_a$ be the scaling dimension of the perturbing operator $\varphi_a$, and let the operator product expansion (OPE) among the perturbing operators be
\begin{equation}
\label{OPE}
\varphi_a \times \varphi_a \rightarrow \sum_{c} C_{abc} \varphi_c.
\end{equation}
Then the one-loop beta functions can be read off:
\begin{equation}
\label{Cardy_beta}
\beta(g_a) = (2 - \Delta_a) g_a - \pi \sum_{b,c} C_{bca} g_b g_c.
\end{equation}

Using the boson correlation function (i = 1,2)
\begin{equation}
\langle \phi_i(\vec{r}) \phi_i(0) \rangle = -\frac{1}{4\pi J_i} \ln(r^2),
\end{equation}
one can identify the scaling dimensions
\begin{equation}
\label{scaling_dim}
\Delta_1 = \Delta_2 = 2; \; \Delta_3 = \frac{1}{\pi}\left(\frac{1}{J_1} + \frac{1}{J_2}, \right)
\end{equation}
and the non-zero OPE coefficients
\begin{equation}
\label{OPEcoefficients}
C_{331} = C_{332} = -\frac{1}{2}; \; C_{133} = -\frac{2}{\pi^2J_1^2}; \; C_{233} = -\frac{2}{\pi^2J_2^2}.
\end{equation}

If we identify the \emph{total} superfluid stiffness $K_1 = J_1 + g_1$ and $K_2 = J_2 + g_2$, then to one-loop order the beta functions follows:
\begin{equation}
\label{betaOneLoop}
\beta(K_1) = \beta(K_2) = \frac{\pi^2}{2} g_3^2; \;
\beta(g_3) = \left[2 - \frac{1}{\pi}\left(\frac{1}{K_1}+\frac{1}{K_2}\right)\right] g_3.
\end{equation}

Clearly $g_3$ is strongly relevant when $K_{1,2}$ are large.  In particular, if the layers are not coupled, the single-layer BKT transition occurs at $K_1 = K_2 = 2/\pi$, and $g_3$ has long become relevant when $K_1$ and $K_2$ are as large.  The implication is that $g_3$ is always a strong perturbation in the low temperature ordered phase, and the validity of a perturbative treatment is questionable at best.

\section{Duality transformation}

The type of duality transformation considered in this work was pioneered by \cite{Jose1977-kb} for the XY model in 2D.  The method can be readily adopted for the 3D XY model (see e.g. \cite{Savit1980-kt}), and also to the bilayer model investigated here.  However, we do find this standard derivation containing a few cryptic steps.  On the other hand, in his attempt to dualize the XY model in 3D \cite{Peskin1978-qq}, Peskin took a longer route that we feel is physically much more transparent.  We will adopt Peskin's treatment to our bilayer model, and flesh out the steps in much greater details that is closer in writing style to lecture notes or a textbook.  We hope the alternative derivation here can serve as useful reference to the broader community.

\subsection{Villain model and the fictitious layer}

We begin with the bilayer XY model with an inter-layer SOJC:
\begin{equation}
\label{XYappendix}
S\! = \!\!\sum_{\substack{i=1,2 \\ r}}\!\!\! -K_i \cos(\vec{\Delta} \phi_i)  - \sum_{r} \lambda \cos(2\phi_1 - 2\phi_2).
\end{equation}
Naively, the equivalent Villain model\cite{Villain1975-ed} is
\begin{equation}
\label{Villain1}
S\! = \sum_{\substack{i=1,2 \\ r}} \! \frac{K_i}{2} \left( \vec{\Delta} \phi_i + 2\pi \vec{m}_i\right)^2
+ \sum_{r} \lambda\left[ (2\phi_1 - 2\phi_2 + 2\pi m_{z})^2 \right].
\end{equation}
But, as we have explained in the main text, it is beneficial to insert a \emph{fictitious layer}, and we instead consider
\begin{equation}
\label{Villain2}
\begin{split}
S\! &= \sum_{\substack{i=1,2 \\ r}} \! \frac{K_i}{2} \left( \vec{\Delta} \phi_i + 2\pi \vec{m}_i\right)^2
+ \sum_{r} \! \frac{K'}{2} \left( \vec{\Delta} \phi' + 2\pi \vec{m}' \right)^2 \\
& \quad + \sum_{r} \lambda\left[ (2\phi_1 - 2\phi' + 2\pi m_{z1})^2 + (2\phi' - 2\phi_2 + 2\pi m_{z2})^2 \right].
\end{split}
\end{equation}
When $K'$ is set to zero, $\phi'$ can be integrated out, yielding exactly \eqref{Villain1}.

We can visualize \eqref{Villain2} as being defined on a 3D cubic lattice, with the $z$-direction truncated to only three layers.  The $\phi$ fields are defined on the lattice sites, and the Villain integers on the links.  We will reserve the vector sign for 2D throughout.  The dual lattice sites are situated at the centers of the cubes.  Therefore the dual lattice will have two 2D layers, and three sets of vertical links.

\subsection{Integer gauge choice and vorticity}

Initially \eqref{Villain2} is in the so-called Villain gauge, where $\phi_1, \phi_2, \phi'$ are compact and take values in $[-\pi, \pi)$, and there is no restriction on the Villain integers.  There is a ``gauge transformation'' that formally allows for the addition of arbitrary integer multiples of $2\pi$ to the local $\phi$ variables: $\phi_1 \rightarrow \phi_1 + 2\pi n_1$, $\phi_1 \rightarrow \phi_2 + 2\pi n_2$ and $\phi' \rightarrow \phi' + 2\pi n'$ if the Villain integers are transformed by:
\begin{equation}
\label{VillainIntegerGauge}
\begin{split}
\vec{m}_1 &\rightarrow \vec{m}_1 - \vec{\Delta} n_1 \\
\vec{m}_2 &\rightarrow \vec{m}_2 - \vec{\Delta} n_2 \\
\vec{m}' &\rightarrow \vec{m}' - \vec{\Delta} n' \\
m_{z1} &\rightarrow m_{z1} - 2(n_1 - n') \\
m_{z2} &\rightarrow m_{z2} - 2(n' - n_2)
\end{split}
\end{equation}
Utilizing this gauge transformation, we can pass to the so-called Peskin gauge, where the $\phi$ variables take arbitrary real values, and the sum over Villain integers are ``gauge-fixed'': two configurations are identified if they are related by \eqref{VillainIntegerGauge}, and the sum is only over non-equivalent configurations.  The different gauge choices represent the same overall degrees of freedom entering the partition function in different ways.

The gauge-fixing requirement on the Villain integers resembles that of the familiar Maxwell theory, with the notable modification that $m_{z1}$ and $m_{z2}$ transform like they are doubly charged.  One expects that the some analogy of the ``curl'' would correctly count the gauge classes.  Thus we define the following integer variables on the plaquettes of the cubic lattice as (indices $\alpha, \beta = x,y$):
\begin{equation}
\label{VillainVorticity}
\begin{split}
w_{1\alpha} &= \epsilon_{\alpha\beta} \left[ 2\Delta_\beta m_{1z} - \left(m_{1\beta} - m'_{\beta} \right)\right] \\
w_{2\alpha} &= \epsilon_{\alpha\beta} \left[ 2\Delta_\beta m_{2z} - \left(m'_{\beta} - m_{2\beta} \right)\right] \\
l_{z1} &= \epsilon_{\alpha\beta} \Delta_\alpha m_{1\beta} \\
l'_{z} &= \epsilon_{\alpha\beta} \Delta_\alpha m'_{\beta} \\
l_{z2} &= \epsilon_{\alpha\beta} \Delta_\alpha m_{2\beta}. \\
\end{split}
\end{equation}
The gauge-fixed sum over $m$ is equivalent to the sum over all integer values of $\vec{w}_1$, $\vec{w}_2$, $l_{z1}$, $l_{z2}$, $l'_{z}$ subjected to the divergence-free constraints imposed on the dual lattice sites:
\begin{equation}
\label{divFreeVorticity}
\vec{\Delta}\cdot\vec{w}_1 + 2(l_{z1} - l'_z) = \vec{\Delta}\cdot\vec{w}_2 + 2(l'_{z} - l_{z2}) = 0
\end{equation}

The new set of integers has the obvious interpretation of \emph{plaquette vorticity}.  The $l$-integers are integer vortices piercing through the corresponding layers.  On the other hand, $\vec{w}_1/2$ and $\vec{w}_2/2$ are the horizontal \emph{half-vortex} lines furrowing in the space between layers, to be identified as the $\mathbb{Z}_2$ domain walls.  These plaquette variables can alternatively be thought of as living on the links of the dual lattice.

The combinations appearing in \eqref{VillainVorticity} will later emerge naturally during the duality transformation.  For now, we will still express everything in terms of the $m$-integers, bearing in mind that their summation is implicitly gauge-fixed.

\subsection{The conserved U(1) current}

We can identify the following components of the U(1) current from \eqref{Villain2}:
\begin{equation}
\label{current}
\begin{split}
\vec{j}_1 &\rightarrow K_1 (\vec{\Delta}\phi_1 + 2\pi \vec{m}_1) \\
\vec{j}_2 &\rightarrow K_2 (\vec{\Delta}\phi_2 + 2\pi \vec{m}_2) \\
\vec{j}' &\rightarrow K' (\vec{\Delta}\phi' + 2\pi \vec{m}') \\
j_{z1} &\rightarrow  4 \lambda (2\phi_1 - 2\phi' + 2\pi m_{z1})\\
j_{z2} &\rightarrow 4 \lambda (2\phi' - 2\phi_2 + 2\pi m_{z2})
\end{split}
\end{equation}
And the equations of motion from \eqref{Villain2} imply current conservation:
\begin{equation}
\vec{\Delta} \cdot \vec{j}_1 + j_{z1} = \vec{\Delta} \cdot \vec{j}_2 - j_{z2}
= \vec{\Delta} \cdot \vec{j}' - j_{z1} + j_{z2} = 0
\end{equation}
Note that when $K'$ is set to zero, $\vec{j}'$ vanishes, and the fictitious layer simply allows tunneling current $j_{z1} = j_{z2}$ to pass through unmolested.

\subsection{Integrating out the $\phi$ fields}

We introduce real-valued auxilliary fields $\vec{h}_1$, $\vec{h}_2$, $\vec{h}'$, $h_{z1}$, $h_{z2}$ on the corresponding lattice links, and perform a Hubbard-Stratonovich transformation on \eqref{Villain2} to get:
\begin{equation}
\label{intermediate1}
\begin{split}
S\! &= \sum_{\text{links}}
\Biggl \lbrace
\frac{\vert h_1\vert^2}{2 K_1}
+ \frac{\vert h_2\vert^2}{2 K_2}
+ \frac{\vert h'\vert^2}{2 K'}
+ \frac{(h_{z1}^2 + h_{z2}^2)}{16\lambda} \\
&\qquad\qquad
- i \big [ 
\frac{\vec{j}_1\cdot \vec{h}_1}{K_1}
+ \frac{\vec{j}_2\cdot \vec{h}_2}{K_2}
+ \frac{\vec{j}'\cdot \vec{h}'}{K'} \\
&\qquad\qquad
+ \frac{\left(h_{z1} j_{z1} + h_{z2} j_{z2} \right)}{8\lambda}
\big ]
\Biggr \rbrace
\end{split}
\end{equation}
The action is now linear in the $\phi$ fields, and they can be integrated out straightforwardly.

Recall that in the Peskin gauge the $\phi$ fields are not compact and take values over the entire real line.  The integral therefore yields Dirac delta functions constraining the real-valued $j$ fields:
\begin{equation}
\label{intermediate2}
\begin{split}
S &\rightarrow  \sum_{\text{links}}
\Biggl \lbrace
\frac{\vert h_1\vert^2}{2 K_1}
+ \frac{\vert h_2\vert^2}{2 K_2}
+ \frac{\vert h'\vert^2}{2 K'}
+ \frac{(h_{z1}^2 + h_{z2}^2)}{16\lambda} \\
&\qquad\qquad
- 2\pi i \big [ 
\vec{m}_1\cdot \vec{h}_1
+ \vec{m}_2\cdot \vec{h}_2
+ \vec{m}'\cdot \vec{h}' \\
&\qquad\qquad
+ \frac{1}{2}\left(h_{z1} m_{z1} + h_{z2} m_{z2} \right)
\big ]
\Biggr \rbrace;
\end{split}
\end{equation}
with the current conservation constraint strictly enforced:
\begin{equation}
\label{hConservation}
\vec{\Delta} \cdot \vec{h}_1 + h_{z1} = \vec{\Delta} \cdot \vec{h}_2 - h_{z2}
= \vec{\Delta} \cdot \vec{h}' - h_{z1} + h_{z2} = 0.
\end{equation}
By the equations of motion, the $h$ fields represent the U(1) current in the dual description.

\subsection{Reparameterization of $h$ and the added (gauge) redundancy}

The divergence-less condition on a vector field is automatically satisfied by writing it as a curl, and we adopt the essence of this idea to rewrite the $h$ fields.  We introduce a set of real-valued variables on the links of the \emph{dual} lattice ($i= 1, 2$ and $\alpha, \beta = x, y$):
\begin{equation}
\label{gauge}
\begin{split}
h_{zi} &= \frac{\epsilon_{\alpha\beta}}{\pi} \Delta_\alpha (\vec{a}_i)_\beta \\
(\vec{h}_1)_{\alpha} &= \frac{\epsilon_{\alpha\beta}}{2\pi} \left( \vec{\Delta} \theta_1 + 2 \vec{a}_1 \right)_\beta \\
(\vec{h}_2)_{\alpha} &= \frac{\epsilon_{\alpha\beta}}{2\pi} \left(\vec{\Delta} \theta_2 - 2\vec{a}_2\right)_\beta \\
(\vec{h}')_{\alpha} &= \frac{\epsilon_{\alpha\beta}}{2\pi} \left( \vec{\Delta} \theta' - 2\vec{a}_1  + 2 \vec{a}_2\right)_\beta \\
\end{split}
\end{equation}
Here $\vec{a}_{1,2}$ lives on the in-plane links of the two layers of the dual lattice, and $\theta_1, \theta', \theta_2$ lives on the $z$-direction dual links.  One can verify that \eqref{hConservation} is trivially satisfied.

This reparameterization introduces redundancy into our description of the model: many configurations of the $\theta$ and $a$ fields lead to the same configuration of the $h$ fields, though a few more steps are needed to beat this redundancy into the familiar form of the U(1) gauge structure.  We make a mental note that these $\theta$ and $a$ fields must be implicitly gauge-fixed, but leave the exact condition vague for now.

\begin{widetext}

Putting \eqref{gauge} into \eqref{intermediate2}, and rearranging the summands as necessary, one transforms the action into
\begin{equation}
\begin{split}
S &\rightarrow \sum_{\text{dual links}} 
\Biggl \lbrace
\frac{\vert \vec{\Delta}\theta_1 + 2\vec{a}_1 \vert^2}{8 \pi^2 K_1}
+ \frac{\vert \vec{\Delta}\theta_2 - 2\vec{a}_2 \vert^2}{8 \pi^2 K_2}
+ \frac{\vert \vec{\Delta}\theta' - 2\vec{a}_1 + 2\vec{a}_2 \vert^2}{8 \pi^2 K'}
+ \frac{(\vec{\Delta}\times\vec{a}_1)^2 + (\vec{\Delta}\times\vec{a}_2)^2)}{16\pi^2\lambda} \\
&\qquad\qquad
- i \left[ 
\theta_1 l_{z1}
+ \theta_2 l_{z2}
+ \theta' l_z'
+ \vec{a}_1 \cdot \vec{w}_1
+ \vec{a}_2 \cdot \vec{w}_2
\right ]
\Biggr \rbrace.
\end{split}
\end{equation}

As promised earlier, the plaquette vorticities $\vec{w}$ and $l$ defined in \eqref{VillainVorticity} make their appearance.  The constraints \eqref{divFreeVorticity} will be enforced by additional auxilliary fields $\chi_1$ and $\chi_2$ living on the dual lattice sites.  The terms added to the action are
\begin{equation}
- i \sum_{\text{dual sites}}
\chi_1 \left[ \vec{\Delta}\cdot\vec{w}_1 + 2(l_{z1} - l'_z) \right]
+ \chi_2 \left[ \vec{\Delta}\cdot\vec{w}_2 + 2(l'_{z} - l_{z2}) \right].
\end{equation}
Since we are setting integers quantities to zero, $\chi_1$ and $\chi_2$ are compact angular fields taking values within $[-\pi, \pi)$.  Again rearrange the summand to expose the vorticities:
\begin{equation}
\label{intermediate3}
\begin{split}
S &\rightarrow \sum_{\text{dual links}} 
\Biggl \lbrace
\frac{\vert \vec{\Delta}B_1 + 2\vec{a}_1 \vert^2}{8 \pi^2 K_1}
+ \frac{\vert \vec{\Delta}B_2 - 2\vec{a}_2 \vert^2}{8 \pi^2 K_2}
+ \frac{\vert \vec{\Delta}B' - 2\vec{a}_1 + 2\vec{a}_2 \vert^2}{8 \pi^2 K'}
+ \frac{(\vec{\Delta}\times\vec{a}_1)^2 + (\vec{\Delta}\times\vec{a}_2)^2)}{16\pi^2\lambda} \\
&\qquad
- i \left[ 
(B_1 + 2 \chi_1) l_{z1}
+ (B_2 - 2\chi_2)l_{z2}
+ (B' - 2\chi_1 + 2\chi_2 )l_z'
+ (\vec{a}_1- \vec{\Delta}\chi_1) \cdot \vec{w}_1
+ (\vec{a}_2 - \vec{\Delta}\chi_1)\cdot \vec{w}_2
\right ]
\Biggr \rbrace.
\end{split}
\end{equation}

\subsection{Cost of vorticity}

We add terms quadratic in vorticities $l$ and $\vec{w}$ to the action by hand.  These are identified as fugacity or chemical potential of the vortex cores.  After all, it is an artifact of the lattice theory that a vortex can be hidden at the center of a plaquette for no cost; in a real material there is an energy cost associated with the suppression of superfluid density at the defect.  These terms are
\begin{equation}
\label{chemicalPotential}
\frac12 \sum_{\text{dual links}} V_1 \vert\vec{w}_1\vert^2 + V_2 \vert\vec{w}_2\vert^2 +  U_1 \, (l_{z1})^2 + U_2 \, (l_{z2})^2 + U' (l'_{z})^2.
\end{equation}
Recall our assertion that vorticity should be separately conserved in each layer.  To prevent the mixing of two layers, $U'$ must be taken to $+\infty$.

The sum over vorticities can be carried out using the Poisson summation formula:
\begin{equation}
\label{Poisson}
\sum_{n=-\infty}^{\infty} \exp\left(-\frac{a}{2} n^2 - i b n \right)
\propto \sum_{k = -\infty}^{\infty} \exp\left(-\frac{(b - 2\pi k)^2}{2a}\right)
\propto \exp\left( \frac{1}{a} \cos(b) \right)
\end{equation}
where we have disregarded overall normalization constants, and by noting that the result of the first step is in the Villain form, we revert the Villain form to its equivalent cosine model.

Applying \eqref{Poisson} to the sum of \eqref{intermediate3} and \eqref{chemicalPotential}, one obtains a fully dualized action:
\begin{equation}
\label{intermediate4}
\begin{split}
S &\rightarrow \sum_{\text{dual links}} 
\Biggl \lbrace
\frac{\vert \vec{\Delta}B_1 + 2\vec{a}_1 \vert^2}{8 \pi^2 K_1}
+ \frac{\vert \vec{\Delta}B_2 - 2\vec{a}_2 \vert^2}{8 \pi^2 K_2}
+ \frac{\vert \vec{\Delta}B' - 2\vec{a}_1 + 2\vec{a}_2 \vert^2}{8 \pi^2 K'}
+ \frac{(\vec{\Delta}\times\vec{a}_1)^2 + (\vec{\Delta}\times\vec{a}_2)^2)}{16\pi^2\lambda} \\
&\qquad\qquad
- \frac{\cos(\vec{\Delta} \chi_1 - \vec{a}_1)}{V_1}
- \frac{\cos(\vec{\Delta} \chi_2 - \vec{a}_2)}{V_2}
- \frac{\cos( B_1 + 2\chi_1)}{U_1}
- \frac{\cos( B_2 - 2\chi_2)}{U_2}
- \frac{\cos( B' - 2\chi_1 + 2\chi_2)}{U'}
\Biggr \rbrace.
\end{split}
\end{equation}

\subsection{Loose ends}

We will take the limit $K' \rightarrow 0^{+}$ and $U' \rightarrow +\infty$ to finally remove the fictitious layer from the model.  The action \eqref{intermediate4} has a truncated-3D version of U(1) gauge symmetry, but to proceed, it is easier to just set $1/U' \rightarrow 0$, and observe that the action now possesses two copies of the familiar U(1) symmetry, one for each dual layer, with $\theta'$ being neutral.

By a U(1) gauge transformation in either of the layers, one can set $\vec{\Delta}B'$ to zero.  And then the limit $K' \rightarrow 0^{+}$ forces $\vec{a}_1 - \vec{a}_2 = 0$.  We shall denote $\vec{a} \equiv \vec{a}_1 = \vec{a}_2$.

We can also compactify the remaining $B_{1,2}$.  Right now they take arbitrary real values, but we may rewrite them as ($i = 1,2$)
\begin{equation}
B_i = \theta_i + 2\pi p_i
\end{equation}
%4
where $\theta_i$ takes value within $[-\pi,\pi)$ and $p_i$ is integer-valued.  The already-compact terms in the action are not affected: $\cos(B_i \pm 2\chi_i) = \cos(\theta_i \pm 2\chi_i)$.  On the other hand, the gradient terms $\vert \vec{\Delta} B_i \pm 2 \vec{a}\vert^2 = \vert \vec{\Delta} \theta_i \pm 2 \vec{a} + 2\pi(\vec\Delta p_i)\vert^2$ now have the Villain form, and we just put them back into the equivalent cosine models likewise.

We have thus arrived at the final form of the lattice dual action (eq (3) in the main text):
\begin{equation}
\label{dual_action}
S_d \! = \sum_r \frac{(\vec{\Delta} \times \vec{a})^2}{8\pi^2\lambda}
	+ \sum_{\substack{i = 1,2 \\ r}} - \frac{\cos(\vec{\Delta}\chi_i - \vec{a})}{V_i}
			- \frac{\cos(2\chi_i - (-1)^i \theta_i)}{U_i}
			- \frac{\cos(\vec{\Delta}\theta_i - (-1)^i 2\vec{a})}{4\pi^2 K_i}
\end{equation}

\end{widetext}

\subsection{Duality dictionary}

Recall that the $h$ fields appearing in the intermediate steps represents the conserved current in the original XY model, and then we represent $h$ using what eventually become $\theta_{1,2}$ and $\vec{a}$.  Retracing our steps, it is not difficult to see that the tunneling current $j_z \equiv j_{z1} = j_{z2}$ is dual to $\vec{\Delta} \times \vec{a}$, while the in-plane currents $\vec{j}_{i}$ are dual to
\begin{equation}
\epsilon_{\alpha\beta} (\vec{j}_{i})_{\beta} \sim -\sin(\Delta_{\alpha} \theta_i - (-1)^u 2a_{\alpha}).
\end{equation} 

Turning off SOJC leaves us with two decoupled XY models, and on the dual side it means the dual gauge field is turned off.  We can then identify $\theta_i$ as the familiar disordered fields from the single-layer duality transformation.

Going back to \eqref{intermediate3}, one sees that $\vec{a}_1$ and $\vec{a}_2$ couples linearly to in-plane vorticities $\vec{w}_1$ and $\vec{w}_2$, respectively.  This implies the identification of the in-plane dual U(1) currents to the in-plane vorticities.  Following the interpretation of \cite{Lee1985-qy}, we identify the ``first-quantized'' world lines of particle excitations created by $e^{\pm i \chi_i}$ as the lines of $\mathbb{Z}_2$ DW.  The in-plane U(1) current is a sort of ``coarse-grained'' description of the DW density.

\bibliography{XY}

\end{document}